# 3D-COMPUTER ANIMATION FOR A YORUBA NATIVE FOLKTALE

Adeyanju I. A., Babalola C. T., Salaudeen K. B. and Oyediran B. D.

Department of Computer Science and Engineering, Ladoke Akintola University of Technology, P.M.B. 4000, Ogbomoso, Nigeria

## ABSTRACT

*Computer graphics has wide range of applications which are implemented into computer animation, computer modeling e.t.c. Since the invention of computer graphics researchers haven't paid much of attentions toward the possibility of converting oral tales otherwise known as folktales into possible cartoon animated videos. This paper is based on how to develop cartoons of local folktales that will be of huge benefits to Nigerians. The activities were divided into 5 stages; analysis, design, development, implementation and evaluation which involved various processes and use of various specialized software and hardware. After the implementation of this project, the video characteristics were evaluated using likert scale. Analysis of 30 user responses indicated that 17 users (56.7%) rated the image quality as excellent, the video and image synchronization was rated as excellent by 9 users (30%), the Background noise was rated excellent by 18 users (60%), the Character Impression was rated Excellent by 11 users (36.67%), the general assessment of the storyline was rated excellent by 17 users (56.7%), the video Impression was rated excellent by 11 users (36.67%) and the voice quality was rated by 10 users (33.33%) as excellent.*

## 1. INTRODUCTION

The advents of computer graphics which are graphics created using computer and the representation of an image data by the computer specifically with the help of specialized graphics hardware and software has had significant impact on many fields like media and has revolutionized animation, movies and the video games industries. Computer graphics has emerged as a field of computer science which studies method for digitally synthesizing and manipulating visual contents [1]. One of the fields of computer graphics is computer animation, which is the use of computers to create animations.

Computer animation is as a technique for creating illusion of movement on a screen, or recording a series of individual states of a dynamic scene [2]. Computer animation can also be viewed as a way to manipulate a sequence of images on a frame by frame basis. The end product of most computer animation is generally known as cartoon. 3D Computer animations create and render digital drawings thereby producing perfect and three dimensional looking animations. Single frame drawings can be designed using standard painting software tools and then composited.

Computer animation uses different techniques to produce animations. Sophisticated mathematics can be used to manipulate complex three dimensional polygons, apply textures, lighting and other effects as well as finally rendering the complete image [3]. This paper focuses on using 3D computer animation techniques to create moving images of a local folktale in the South West (Yoruba speaking) of Nigeria. It is hoped that the animation of local folktales will help children to learn more about their cultural heritage and reduce the influence of foreign cartoons. In a modern world where computer and technology has revolutionized everything, culture which folktales is part of, should not be left behind. Especially, in Nigeria where most of our cultural practices have been pushed away, gone are the days where moonlight stories, fairy tales and some other interesting cultural stories were been told. This paper aims at reversing this trend  using computer animation to produce a cartoon of a selected folktale will help  to revive local cultural practice where lessons can be learnt similar to when it was narrated orally before the advent of Computers.

## 2. RELATED WORK

The impact of computer animation on children and their behaviors has previously been investigated [4]. The focus was on violence and formation of characters in children and how animations create ideas in children in comparison with other media. The content and characteristics of 2D and 3D animation on children in visual media was studied especially in TV, Internet and Film in Kerala part of India [4].

Similarly, [5] reported that computer animations accompanied with traditional teaching increases the performance of high school biology students. The teaching of chemical bonding using the animation and jigsaw techniques was found to be more effective than the traditional teaching methods [6]. Other similar studies reported that animations improved students' academic performance in mathematics and statistics [7] [8].

The revolution in information and communication technologies provided a motivation for the emergence of the new media [9]. In this context, new media are regarded as the means of information transmission and dissemination that surpass the oral and the written media such as internet, mobile telephony, television and other forms of digital communication.

Cartoons are important features of newspapers and magazines with a lot of readers following such cartoon series. The effectiveness of cartoons in print media was examined [10]. The authors also looked at the safety such cartoons offer in terms of freedom of expression, especially during the intolerant military regimes [10].

An animated folk tales edutainment software was developed to motivate socio-cultural awareness among children and adolescents [11]. The software included animated cartoons, movies and digital games in Malay to motivate socio-cultural awareness. The effectiveness of exposure to news, cartoons, and films as three different types of authentic audiovisual programs was analyzed as a means of improving the language proficiency of learners [12].  Their results showed that audiovisual programs generally are a great source of language input for teaching purposes. Animations with good story lines seem to motivate the learners to better understand the language [12].

## 3. FOLKTALE ANIMATION DESIGN AND IMPLEMENTATION

The activities in the design of an animation can be subdivided into six main parts as shown in Fig 1. These stages are standard in computer animation. We therefore discuss the design and implementation of our animation with respect to these stages.

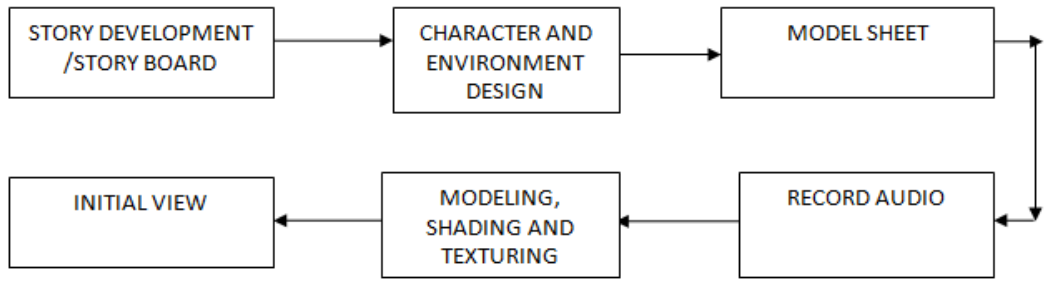

Fig 1: Flow Diagram of the design of a computer animation

## 3.1    Story Development

This is the first step for any form of animation creation; without a story there is no animation. The story is developed in a form of a script which contains dialogue for voiceover and other information that will help in developing the storyboard. Yoruba Folktales were studied and compared to identify for socio-cultural values. The story "Ijapa ati Atioro" (the tortoise and the bird) was selected due to its simplicity and important lessons that can learnt from the story.

The setting of the story is in a village based on the original story line, a script was written with dialogues for both characters in the story. The story started with a scene showing one of the characters waking up from his sleep hungry with nothing to eat as there was famine in the village. As the story progresses, the second character is revealed with a scene of how healthy and happy he is living despite the famine because he is able to fly to a far distance to obtain food. The ending of the story is narrated by a voice to ensure that the lessons that can be learnt from the story are understood by the audience. The moral of the story includes hard work and contentment. The script contains dialogue and the information that will help in the development of the story board.

## 3.2    Storyboard

It is made of sequence of drawing of each scene in the form of a comic strip. This is the most important aspect in creating any animation. It gives the director, modeler and animator the foreknowledge on how the scenes are going to look like. This process ensure that the story have been thought through and to give the general overview of the story. This stage also gives room for necessary revisions such as adding or removing from the story before going on with the other processes. Sequence of drawing of each scene in the form of a comic strip is done directly using Adobe Photoshop with the aid of a (Wacom) Tablet connected to the computer via a USB port. The basic tool used in the Photoshop is the brush which functions like the pencil. Every strip in the storyboard was drawn on different layers.

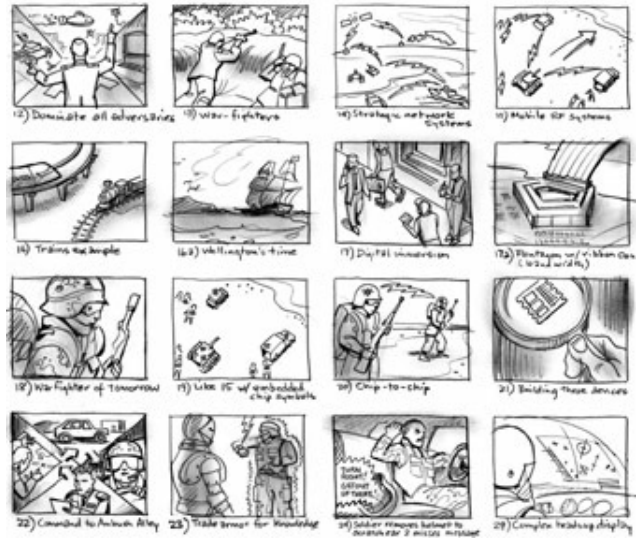

Fig 2: An Example of a Story Board (Source: Wikipedia, 2014)

### 3.3    Character and Environment Layout

The layout and character design is the next stage after the story board. During this process, the environment and characters in the story are designed based on the drawings made during the previous (story boarding) stage. This is also the time to consider the color of choice for the design. The story has only two characters which are Ijapa and Atioro. These characters were designed based on local identity in terms of physical appearance, attires and styles. The environment and characters in the story are designed based on the drawings made during the storyboarding stage. Figure 4 and 5 are illustrations of the characters while Figure 6 shows a traditional setting of the environment.

The characters were firstly sketched on a paper, and then all characters were drawn directly in Adobe Photoshop as shown in Fig 3.

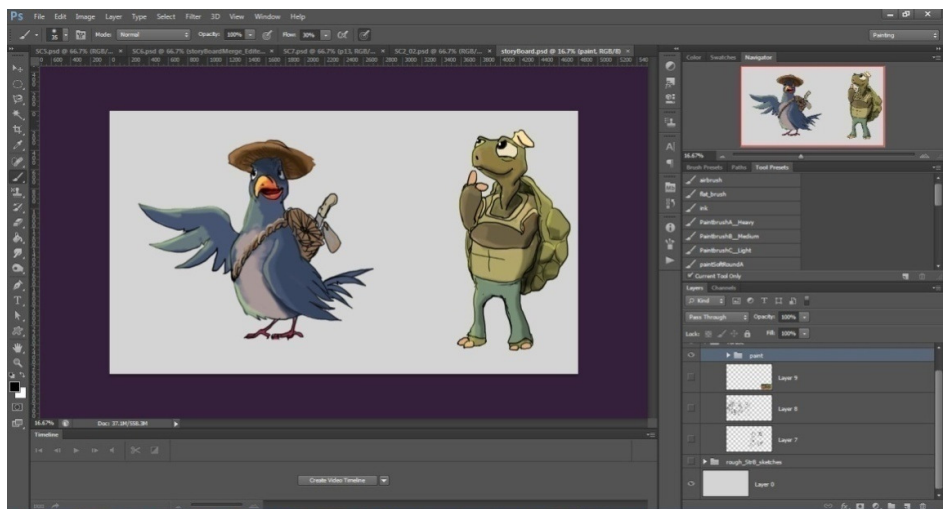

Fig. 3: A Snapshot of the Modeling of the Character in Adobe Photoshop

## 3.4 Model Sheet

Characters will always have peculiar traits that will reflect in both their visual appearance and behavior. Everything that makes a character unique is contained in its model sheet – starting with the body structure, its costumes, poses and facial expressions. In our animation, there are two characters. The first character is Ijapa (Tortoise), a covetous, deceptive, cunning and very lazy animal who is slow-moving. Ijapa is also a land-dwelling reptile with a large dome-shaped shell into which it can retract its head and limbs. On the other hand, Atioro (a bird) is a very hardworking, agile, active and responsible two-legged, warm-blooded animal with wings. Atioro also has a beak and a body covered with feathers.

## 3.5 Animatics

This is a composed form of the storyboard created to illustrate mainly the timing, visual effects and the camera movements together with the soundtrack and voiceover. Animatic is a quick representation of the animation with the aim of showing the timing and sound. We did our animatics in Adobe Photoshop extended software which has the capability to show time frames. The animatic was done in the timeline panel with an arrangement of pictures in frames. There were 24 frames per second and required changes in the timing were effected before the real animation.

## 3.6 3D Modeling

The process of modeling in animation enables conversion of a 2D piece of concept art into a 3D model. Our modeling was done using a sculpting tool (z-brush) since it is known to be better for organic modeling of objects without definite shape. A base mesh was created using a z-sphere and sculpting brushes, and then adjusted into forms and shapes needed to form the character model. The model was then saved and exported to Maya for re-topology. Re-topology is the process of reducing the level of the polygons in the body and making the model animated. Different stages involved in the 3D modeling of Ijapa are shown in Figure 4.

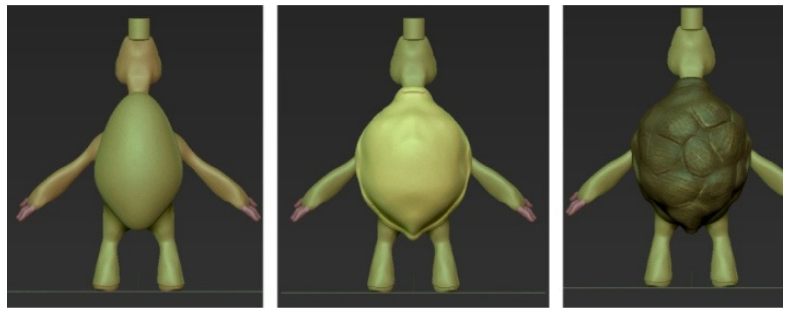

Fig. 4: A Snapshot of the stages in the Modeling the character "Ijapa"

## 3.7 Rigging

Rigging is the process of adding bones to a character or defining the movement of a mechanical object. It is central to the animation process as it will show how a character appears when deformed into different poses. The rigging process involves creating the skeleton, adding skin to the skeletal part of the body and then attaching the different parts together. Using Maya software, the creation of skeleton is done by selecting animation on the drop down menu, from the skeleton menu and clicking on joint tool followed by clicking on the part in which the skeleton is required.

Skin can be added by clicking on smooth bind on the skin menu, to fix the skin and skeleton together. Fig 5 shows a snapshot of character "Atioro" being rigged.

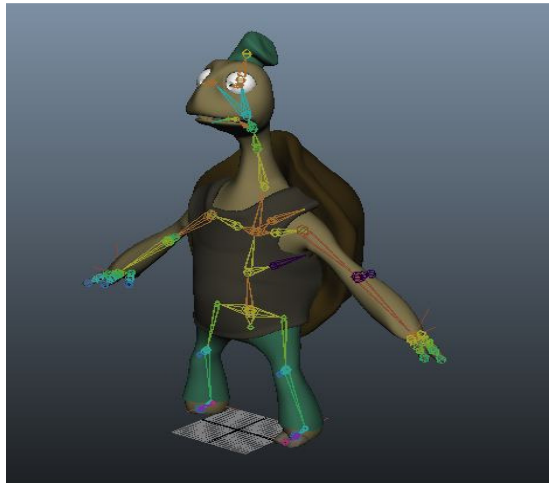

Fig. 5: A Snapshot of the Rigging Process

### 3.8  Shading & Texturing

In this phase, materials, textures, and colors are added to the models created. Every component of the model receives a different Shader; which is the material that gives it an appropriate look. In our work, Z-brush software was used for shading which comes before texturing. Mapping is done under texturing to place texture in a particular position in the character to be designed.

### 3.9  Lighting

3D scenes become alive with digital lights placed to illuminate the created models. This is similar to the use of lighting rigs on a movie set to illuminate actors and actresses. This stage includes placing lights, defining light properties and defining how light interacts with different types of materials. It also involve determining the qualities and complexities of the realistic textures involved, how the position and intensity of lights affect mood and believability, as well as color theory and harmony. These are required to establish direct and reflected lighting and shadows for each assigned shot, ensuring that each shot fits within the continuity of a sequence.

In our work, Maya software was used. Maya has a default lighting which that allows the scene to be viewed clearly. We also created an exterior lighting and ambient interior lighting using the same Maya software.

### 3.10 Animation

This is the phase where characters are given "life" and motion.  In  the  animation  process,  we created a new document for a particular scene. We imported the characters and the environment into the scene created. The cameras were set based on what we have in the animatics and we then started the animation using the animatics as our guideline. Each character is keyed into the required poses and positions. Sound was also imported using the trax editor sub-menu under the animation editor menu of software.

### 3.11 Rendering

In this phase, the shadows, reflections, ambient occlusion are rendered out as different passes for later compositing. Also special effects such depth of field, motion blurs, fog, smoke were integrated into the scene for rendering. Rendering is the process of converting the animation into video or sequential pictures for compositing purposes. The video format for our rendering was AVI and in case of sequential pictures, the format is JPEG.

### 3.12 Compositing

The is the final stage in which all the rendered image sequences are combined together to outputted as the designed video format. Some minor edits were made in terms of brightness, contrast and color correction. Movie introduction and the title were also included at this stage. Compositing is done in "Aftereffect" software by importing the rendered video format into the software and carrying out editing such as color correction, importing background for the scene among others.

### 3.13 Initial View

This is a special view that is shown before the rest of the application is loaded. The emotional influence that is conveyed is important and is capable of developing thinking skills to the viewer of any of the stories presented. Fig 6 and 7 show two initial views of our animation after compositing.

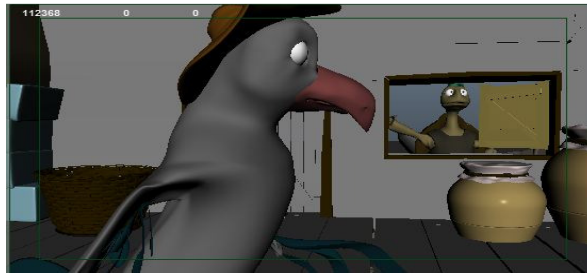

Fig. 6: A snapshot of Ijapa and Atioro in Atioro's House

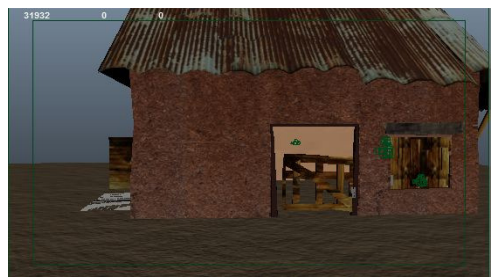

Fig. 7: A snapshot of Ijapa's House

## 4. Evaluation Results

A likert-scale based questionnaire was designed to evaluate our developed animation. The questionnaire was used to get the required information from people about different aspects of the computer animation including image quality, voice and image synchronization, background noise, character impression of the video, voice quality and general assessment of the storyline. Thirty participants were requested to watch the animation and thereafter complete the questionnaire. Therefore, thirty user responses were received; the results of data analysis of individual Likert items are presented in the Table 1 below.

Table 1: Evaluation Results

| Criteria | Excellent (5) | Very good (4) | Good (3) | Fair (2) | Poor (1) |
|---|---|---|---|---|---|
| Image quality | 56.7% | 43.3% | 0% | 0% | 0% |
| Voice and image synchronization | 30% | 36.67% | 23.33% | 10% | 0% |
| Background noise | 60% | 30% | 3.33% | 6.67% | 0% |
| Character impression | 36.67% | 53.33% | 10% | 0% | 0% |
| General assessment of the storyline | 56.7% | 33.3% | 6.67% | 3.33% | 0% |
| Impression of the video | 36.67% | 50% | 10% | 3.33% | 0% |
| Voice quality | 33.3% | 56.7% | 6.67% | 3.33% | 0% |

Table 1 indicates that the Image quality was highly rated with all participants rating the animation as excellent or very good. Similarly, participants liked the character impression with no user rating it as fair or poor. The other five (voice and image synchronization, background noise, video impression, voice quality and general assessment) likert criteria had a few users rating them as fair; this shows that our animation can be improved further in these areas. However, none of the users rated any criterion as poor indicating that our animation is above average.

## 5. CONCLUSION

People tend to better understand any message being sent through visual means than those sent through text or sound because not everyone can read or hear but everyone can understand and interpret whatever they see. Therefore, there is the need to develop animated cartoons of native local folktales which helps to pass lessons across to the audience as well as improving the local cultural heritage. In this work, Maya Autodesk, Adobe Photoshop extended, Zbrush and Adobe Aftereffect CS6 Software were used in the development of a Yoruba native folktale animation. User evaluation indicated that the develop animation satisfies some basic criteria expected of such cartoon. We intend to repeat the animation development for other local folktales and extend our work to cover edutainment software applications for people to learn more from their local culture.

**AUTHORS' BIOOGRAPHY**

Dr. (Engr) Ibrahim Adepoju ADEYANJU is a Lecturer at the Department of Computer Science and Engineering, Ladoke Akintola University of Technology, Ogbomoso, Nigeria where he teaches undergraduate and postgraduate students, supervises students' project/thesis and does research in the area of Intelligent Systems, Machine Learning, Pattern Recognition and Artificial Intelligence.  Ibrahim obtained a Bachelor degree (B.Tech. with Honours) in Computer Engineering from Ladoke Akintola University of Technology, Ogbomoso, Oyo state, Nigeria and his Masters (MSc.) degree in Computing Information Engineering and Doctorate (PhD) degree from the Robert Gordon University, Aberdeen, United Kingdom. Dr. Adeyanju has several peer reviewed journal articles and conference papers across different research areas of Computer Science and Engineering including Pattern Recognition, Distributed Constraints Satisfaction, Artificial Intelligence, Information Retrieval, Information Systems and Machine Learning. Ibrahim is happily married with kids and enjoys reading during his leisure times.

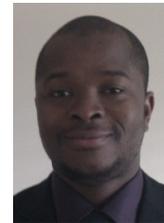

Other authors (Babalola C. T., Salaudeen K. B., Oyediran B. D.) are students who just completed their undergraduate studies at the Ladoke Akintola University of Technology, Ogbomoso, Nigeria. The work presented in this paper was part of the Bachelor's project in Computer Engineering.